# Near-optimal demand side management in retail electricity markets with coupling constraints via indirect mechanism design

G. Tsaousoglou, K. Steriotis, N. Efthymiopoulos, K. Smpoukis, E. Varvarigos

*Abstract--* **Recently there have been several historical changes in electricity networks that necessitate the development of Demand Side Management (DSM). The main objective of DSM is to achieve an aggregated consumption pattern that is efficient in terms of energy cost reduction, welfare maximization and/or satisfaction of network constraints. This is generally envisaged by encouraging electricity use at low-peak times. In this paper, we consider a system with strategic, price-anticipating consumers with private preferences that choose their electricity consumption patterns so as to maximize their own benefit. In this context, we take on the problem of coordinating the consumers' consumption behavior without sacrificing their welfare (Quality of Experience). In order to tackle this problem, we draw on concepts of indirect mechanism design and propose a DSM architecture that is able to fulfill specific system-wide constraints (e.g. energy cost reduction) and simultaneously achieve welfare that is very close to optimal. The proposed billing rule preserves both the budget-balance and the individual rationality properties. According to our evaluation, the proposed DSM architecture achieves a close to optimal allocation (1%-3% gap), compared to an "optimal" system that would use central optimization of user loads without user consensus or protection of their privacy.**

*Index Terms*--Demand Side Management, smart grid, mechanism design, constraint satisfaction.

## I. INTRODUCTION

Modern electricity networks face a number of important developments and subsequent challenges. These are technical (e.g. supply-demand balance and network stability upon high RES penetration), financial (e.g. market liberalization, bottom-up financing) and political (e.g. democratization of energy production systems and energy independency of communities and countries). Energy utilization and energy efficiency is at the center of attention in such systems. In particular, Demand Side Management (DSM) refers to the idea of driving electricity consumption at times where it is more efficient to serve it. As analyzed in [1], residential participation in DSM is commonly envisaged via aggregated participation because of implementation and scalability issues. An Electricity Service Provider (ESP) is considered for the role of aggregating and coordinating the users' actions. Applying direct control over the end-users' loads is not an attractive option since it comes with massive consumer dissatisfaction and arbitrary load prioritization, which leads to loss of social welfare. Along with the trend of liberalization of the electricity market, principles of economics that are already applied in most markets are now becoming more relevant to the electricity market as well. Thus, the state of the art approach to DSM is to motivate electricity consumers towards economically efficient consumption patterns by providing monetary incentives. That is, consumers are expected to modify their consumption patterns voluntarily in response to pricing signals.

Nevertheless, each user is typically trying to optimize his/her own objective, which may or may not be in line with the social objective. A particular stream of game theory called mechanism design is essentially the tool for designing rules (namely, an allocation rule through which end users determine their consumption pattern, and a billing rule through which their bills are determined) for systems with strategic participants holding private information, such that the system at equilibrium has good performance guarantees.

Modern ESPs in the era of the smart grid have to embed DSM in their business models. A DSM architecture includes the mechanism (allocation rule and billing rule) through which the DSM participants (namely, the users and the ESP) interact as well as the local algorithm through which each participant decided his/her actions. Through a carefully designed DSM architecture, we can hopefully bring the system to an efficient state, even though the designer does not directly control the decision variables. According to our requirement analysis [2], a DSM architecture has to fulfill three properties, described in the following subsections.

### A. Welfare Maximization

The first property is the maximization of the welfare (i.e. the aggregated users' utility). The utility of a user/energy consumer is defined as the difference between: i) a metric (noted here as valuation function) that quantifies how much the user valuates/appreciates a specific energy consumption profile/pattern and ii) the bill that the user has to pay for it.

Maximizing the welfare through mechanism design can be relatively easy or really challenging, depending on the assumptions made about the actual users' behavior and

This work received funding from the European Union's Horizon 2020 research and innovation programme under grant agreement No. 731767 in the context of the SOCIALENERGY project.

G. Tsaousoglou, K. Steriotis, N. Efthymiopoulos, K. Smpoukis are with the National Technical University of Athens (NTUA), Greece. E. Varvarigos is with Monash University, Melbourne, Australia.

preferences. Making strong assumptions on the form of user's preferences makes the system conducive to theoretically strong results but the validity of these assumptions is often questionable [3]. Also, a common assumption regarding the user's behavior refers to the user being modeled as a price-taker, which means not considering the effect of his/her own decisions on the electricity price. While this might be relevant for large systems, in emerging energy communities and decentralized systems this assumption no longer holds and the user might be a price-anticipator. The latter user model only makes things more complicated when it comes to welfare maximization and it is avoided in most of the literature (see [4] and references therein). In contrast, in the present paper users are price anticipators.

Finally, the aggregated users' utility alone is not enough. A typically desired property is the property of individual rationality. A mechanism is called individually rational if each and every user benefits from participating in it. In other words, at equilibrium, each user is better-off participating in the DSM, rather than not participating.

### B. Budget-balance

We consider a benevolent ESP that acts on behalf of the users and not against them. The ESP is not a profit maximizing entity but a representative of the users and their interests. Budget-balance refers to the fact that the mechanism is not required to subsidize the DSM participation nor does it extract a surplus from the users, but only divides the system's energy cost among users. Indicative use cases of this business model are: i) the case of energy cooperatives [5], ii) public companies [6] around public authorities acting as ESPs, iii) private monopolistic companies with regulated profit margins, iv) virtual associations of users [7] v) islanded energy communities and vi) any other use case in which the ESP's primary interest is the welfare of the users in its portfolio.

### C. Constraint Satisfaction

The third property is the coordination of the aggregated users' consumption in order to satisfy system-wide constraints. Such constraints indicatively aim to

Case a) keep the aggregated consumption below a certain threshold at all times or

Case b) keep the system's overall cost within certain margins.

The necessity of satisfying such constraints is met in many use cases in modern smart grids which include:
1) Enhancing the self-sufficiency of the community
2) Keeping islanded microgrids economically viable [8]
3) Mitigate suppliers' exercise of market power by taking coordinated action to reduce the demand in the face of such situations [9]
4) Meeting the physical network's constraints by implementing the DSO's orders
5) Enhancing the community's participation in flexibility markets [10, 11]
6) Reducing $CO_2$ emissions and respecting modern legal frameworks towards energy cost reduction [12]
7) Enhancing RES penetration by adapting demand to the intermittent generation [13]

From a technical point of view, satisfying a system-wide constraint can be a challenge. In particular, constraint satisfaction typically depends on the aggregated consumption profile of end users. This couples the system's decision variables that are controlled by different users, which brings a fair amount of complications in the underlying n-person game [14]. The proposed DSM architecture can be used for both cases a) and b) of constraints described above. In this paper, we present a theoretical analysis for case a), which is the most difficult of the two but in the evaluation section we present simulations for both cases.

Further requirements might apply depending on the context and the particular business model of the system. Designing a DSM architecture that exhibits specific properties tailored to each specific business model is an open research topic. In the context of SOCIALENERGY [15] we design allocation and billing rules for various settings, including real-time consumption curtailments ([16], [17]) and day-ahead consumption scheduling ([18] and the present paper). Finally, various studies have proposed solutions for aligning the user's day-ahead consumption schedule with his/her updated real-time needs ([19]-[22]).

Summarizing the above, the contribution of this paper is the design of a DSM architecture that is able to meet system-wide constraints (e.g. energy cost reduction) and at the same time achieve users' welfare very close to optimal. The proposed scheme also preserves both the budget-balance and the individual rationality properties.

## II. RELATED WORK

In the DSM context described above, we set three main requirements for the proposed mechanism. We need a DSM architecture that: a) achieves close to optimal users' welfare, b) preserves the *budget-balance* property, and c) provides the ESP with the ability to control the overall consumption cost (satisfaction of a system constraint).

The welfare-maximizing requirement is highly dependent on user modeling. That is, a theoretically optimal allocation can be achieved, only under certain assumptions on the users' preferences representation. DSM studies can be categorized into three main branches with respect to how they model user preferences.

The first branch includes many works (e.g. [22]-[32]) that consider users who exhibit no preference towards the consumption pattern, as long as their whole load is satisfied within a defined time interval. In simple words, users set constraints on their consumption but are no preferences among the time intervals as long as the consumption constraints are met.

The second branch of the literature (e.g. [1] and [33]-[39]), considers user preferences and price sensitive consumption patterns. The study in [33], approaches the solution with a regret-based algorithm, [35] with Simulated Annealing, and the rest of the works typically formulate a convex optimization problem and reach the optimal solution by solving its dual problem. During this process, the ESP and the users solve their local problems and exchange messages. Under the assumption of price-taking users, the final allocation is welfare-maximizing.

In the third branch (e.g. [4], [40]) this assumption has been relaxed and users are considered as price-anticipators, that is, they consider the effect of their actions on the prices. In this case, the dual approach no longer achieves welfare maximization, as analyzed in [41, chapter 21]. So, the studies in this third branch opt for a Vickrey-Clarke-Groves (VCG) mechanism. However, the practical applicability of the VCG mechanism is highly debated because it is a direct mechanism (requires users to reveal their preferences to the ESP), which raises not only privacy but also representation issues (see [19] for a more detailed analysis).

From the above three user model research branches, only the first one preserves the *budget-balance* property. The convex optimization approach typically ends up with the market-clearing prices and extracts a big surplus from the users, especially when the latter are price-takers. Also, the VCG mechanism is inherently not *budget-balanced*.

Finally, constraint satisfaction complicates things when it comes to indirect mechanisms. This is because typical market-clearing approaches are often not suitable for constraint satisfaction, especially when the constraints couple the optimization variables. Thus, the works that induce some kind of controllability, either relax the welfare-maximization requirement [42], or the user preferences modeling [43], or adopt a central optimization approach [44], [45] with a consequent assumption of direct control on user loads. Also, dual optimization approaches can apply some control on the consumption patterns by manipulating the prices, but that comes at the expense of high user dissatisfaction.

In this work we present a DSM architecture for price-anticipating users that: i) achieves near optimal welfare (reaches 91%-99% of the optimal value), ii) is theoretically proven to preserve the *budget balance* and the *individual rationality* properties, iii) provides the ESP with controllability over the overall system's cost (which is a coupling and quadratic constraint). To the best of our knowledge, this is the first work to satisfy all three of the requirements described above.

## III. SYSTEM MODEL

We consider an electricity market comprised of an Electricity Service Provider (ESP) and a set $N \triangleq \{1,2,\dots,n\}$ of self-interested consumers, hereinafter referred to as users. We also consider a discrete representation of time, where continuous time is divided into timeslots $t \in H$, of equal duration $s$, where $H \triangleq \{1,2,\dots,m\}$ represents the scheduling horizon. A user possesses a number of controllable appliances where each appliance bears an energy demand. We consider each appliance as one user, for ease of presentation and without loss of generality. Thus, we will use the terms "user" and "appliance" interchangeably throughout.

### User & Appliance modeling

An appliance $i$ requires an amount of energy for operation. For example, if an appliance's operating power is 1W, and $s = 1h$, then the energy that the appliance consumes in one timeslot of operation is $1Wh$. This energy consumption is controllable via a decision variable $x_i^t$, which denotes the amount of energy consumed by appliance $i \in N$, at timeslot $t \in H$. Throughout this paper we assume $x_i^t \geq 0$. Each appliance $i$ is characterized by

i) a feasible consumption set, defined by a set of constraints on $x_i^t$, which is presented below and

ii) a valuation function of the energy that $i$ consumes throughout $H$.

The aforementioned set of constraints includes upper and lower consumption bounds, restrictions on consumption timeslots and a coupling constraint. More specifically, appliance $i$ cannot consume more than an upper bound $\overline{x_i}$, that is,

$$0 \leq x_i^t \leq \overline{x_i} \quad (1)$$

An appliance $i$ also bears a set of timeslots $h_i \subseteq H$, in which its operation is feasible (e.g., an electric vehicle can be plugged in only at timeslots during which its owner is home):

$$x_i^t = 0, \quad t \notin h_i \quad (2)$$

We denote an appliance's feasible consumption profile, as a vector $\boldsymbol{x_i} = \{x_i^1, x_i^2, \dots, x_i^m\} \in \mathcal{X}_i$, where $x_i^t$ satisfies (1), (2) and $\mathcal{X}_i \subseteq \mathbb{R}^m$ denotes the feasible set for $i$'s consumption profile:

$$\mathcal{X}_i \triangleq \{\boldsymbol{x_i} \mid x_i^t \text{ such that } (1), (2) \text{ hold}\}, i \in N \quad (3)$$

Finally, the $n \times m$ matrix containing all users' consumptions at all timeslots is denoted as $X = \{\boldsymbol{x_1}, \boldsymbol{x_2}, \dots, \boldsymbol{x_n}\} \in \mathcal{X}$ where $\mathcal{X} = \{\mathcal{X}_i\}_{i \in N}$ denotes the Cartesian product of the $\mathcal{X}_i$'s.

The valuation function is expressed in monetary units (\$), and it is private (the user does not share it with the ESP or other users). It is generally a function of $\boldsymbol{x_i}$ and expresses the maximum amount of money that a user is willing to pay for the operation profile $\boldsymbol{x_i}$. The valuation function $v_i(\boldsymbol{x_i})$ can take various forms, depending on the appliance. Let $\boldsymbol{0}^m$ denote the $m$-vector with all of its elements equal to zero. We adopt some common assumptions ([1], [4], [34]-[39]) based on microeconomics theory [46] on the form of $v_i(\boldsymbol{x_i})$:

*Assumption 1:* Zero consumption brings zero value to the user:

$$v_i(\boldsymbol{0}^m) = 0 \quad (4a)$$

*Assumption 2:* Consuming more does not make the user less happy. That is, for two arbitrary vectors $\boldsymbol{x_{iA}}, \boldsymbol{x_{iB}}$, we have:

$$v_i(\boldsymbol{x_{iA}}) \leq v_i(\boldsymbol{x_{iA}} + \boldsymbol{x_{iB}}), \quad \forall \boldsymbol{x_{iA}}, \boldsymbol{x_{iB}} \quad (4b)$$

*Assumption 3: (concavity)* for two arbitrary vectors $\boldsymbol{x_{iA}}, \boldsymbol{x_{iB}}$ and for any scalar $0 < a < 1$:

$$av_i(\boldsymbol{x_{iA}}) + (1-a)v_i(\boldsymbol{x_{iB}}) \leq v_i(a \cdot \boldsymbol{x_{iA}} + (1-a) \cdot \boldsymbol{x_{iB}}) \quad (4c)$$

Finally, the user's utility is defined as the difference between the user's valuation for his/her consumption profile and the bill he/she has to pay for it:

$$U_i(\boldsymbol{x_i}) = v_i(\boldsymbol{x_i}) - b_i(\boldsymbol{x_i}) \quad (5)$$

### System Cost & Electricity Billing

The ESP is responsible for purchasing energy from the grid and delivering it to the users. We assume that the ESP faces a per-timeslot cost that is a strictly increasing function $C^t(\cdot)$ of the aggregated consumption $\sum_{i \in N} x_i^t$. In particular, quadratic or piecewise linear functions are widely used in the literature [22]-[39], to model the generation cost of marginal units as explained in [47]. We present the case for quadratic cost:

$$C^t(\textstyle\sum_{i\in N} x_i^t) = c \cdot (\sum_{i\in N} x_i^t)^2 \qquad (6)$$

As explained in the introduction, we consider a use case where the ESP needs to be able to control the system's cost so as to keep it below a certain threshold $C_{ref}$. Moreover, we consider a benevolent ESP that acts on behalf of the users and not against them. We assume that, for the scheduling horizon, the ESP collects the financial cost $C$ which is:

$$C = \textstyle\sum_{t\in H} C^t(\sum_{i\in N} x_i^t) \qquad (7)$$

by applying a billing rule $b(x_i)$ to each user. We state some requirements for $b(x_i)$:

*Requirement 1:* The sum of the users' bills should add up to the system's cost:

$$\textstyle\sum_{i\in N} b(x_i) = C^t(\sum_{i\in N} x_i^t) \qquad (8)$$

Eq. (8) captures the *budget balance* property analyzed in the introduction.

*Requirement 2:* At equilibrium, each user should have weakly positive utility.

This is equivalent to stating that each user should be better-off participating in the mechanism rather than not participating. This is equivalent to the *individual rationality* property.

**ESP-user interaction & implementation**

We assume a communication network, built on top of the power grid, allowing the ESP and the users to exchange messages. In particular, in order for an indirect mechanism to be implemented, we assume that the users can respond to demand queries. That is, the ESP provides the user with the necessary billing data and the user is expected to respond with his/her demand, that is, with the desired consumption vector $x_i$ that maximizes the user's utility $U_i(x_i)$ given from Eq. (5).

Since an efficient allocation involves a certain degree of coordination among users, it may take a number of message exchanges between the ESP and each user to converge to equilibrium. For this reason, we expect the user to respond to each demand query in a reasonable amount of time. A commonly accepted response time in computer science is a time that is, in the worst case, polynomial in bits of precision required. For the latter property to hold, the billing rule should be simple enough. A sufficient condition that fulfills this property is captured in a third requirement, which is:

*Requirement 3*: The user's bill $b(x_i)$, is convex in $x_i$.

To justify the sufficiency of *Requirement 3,* recall the definition of the user's utility from Eq. (5). The first term is concave by *Assumption 3*. A convex $b(x_i)$ makes the user's utility $U_i(x_i)$ concave in $x_i$. Thus, the user's response to a demand query becomes a convex optimization problem, which is tractable.

## IV. PROBLEM FORMULATION

In this section, we formalize the problem to be solved, which is maximizing the aggregated users' utility (Eq. 9):

$$\max_{x_i \in X_i, i\in N} \textstyle\sum_{i\in N}(v_i(x_i) - b(x_i)) \qquad (9)$$

while keeping the system's cost below a predefined threshold $C_{ref}$. By using Eq. (8) in Eq. (9) we have:

$$\max_{x_i\in X_i, i\in N}\{\textstyle\sum_{i\in N}[v_i(x_i)] - \sum_{t\in H}[C^t(\sum_{i\in N} x_i^t)]\} \quad (10)$$
$$s.t. \ \textstyle\sum_{t\in H}[C^t(\sum_{i\in N} x_i^t)] \le C_{ref}, \qquad (10a)$$

Constraint (10a) couples the variables $x_i^t$ across both $i \in N$ and $t \in T$. We will demonstrate that this is a standard convex optimization problem where a concave function is maximized over a convex set $\mathcal{X} \subseteq \mathbb{R}^{n\times m}$ that is defined by the inequality constraints (1), (10a) and the equality constraint (2).

*Lemma 1:* The problem defined by Eq. (10) under constraints (1), (2) and (10a) is a convex optimization problem. In particular:

i) The objective function $f(X) = \sum_{i\in N}[v_i(x_i)] - \sum_{t\in H}[C^t(\sum_{i\in N} x_i^t)]$ is concave in $X \in \mathcal{X}$.

ii) Inequality constraint functions (1), (10a) are convex in $X \in \mathcal{X}$

iii) Equality constraint functions (2) are affine in $X \in \mathcal{X}$

*Proof*:

i) Since $\sum_{i\in N}[v_i(x_i)]$ is a sum of concave functions in subspaces of $\mathcal{X}$, it is concave in $\mathcal{X}$. Let $\mathbf{1}_n$ be the all-ones n-dimensional vector and $\mathbf{1}_{n\times n}$ the all-ones $n \times n$ dimensional matrix. Let also $x^t \triangleq (x_1^t, x_2^t, \dots, x_n^t)^T$ be the vector containing all the users' consumptions in timeslot $t$. Then

$$C^t(\textstyle\sum_{i\in N} x_i^t) = c \cdot ((\mathbf{1}_n)^T \cdot x^t)^2 = c \cdot (x^t)^T \mathbf{1}_{n\times n} x^t$$

is convex because it is a quadratic function and $\mathbf{1}_{n\times n}$ is positive semi-definite. Therefore, $-\sum_{t\in H}[C^t(\sum_{i\in N} x_i^t)]$ is concave in $\mathcal{X}$, as a sum of concave functions in subspaces of $\mathcal{X}$ and (i) is true because $f$ is a sum of concave functions.

ii) Constraint (1) is trivially convex and (10a) is also convex as shown in the second term of $f$

iii) Constraint (2) is trivially affine, for all $i \in N$, in a subspace of $\mathcal{X}$, and so it is also affine in $\mathcal{X}$. ∎

Thus, problem (10) is convex and has a global optimal solution. If valuations $v_i(x_i) \triangleq v_i(x_i^t; t \in H)$ were known, it could be solved through the use of an interior point method. However, $v_i(x_i)$ is private. Moreover, we assume strategic users who opt for maximizing their own utility, that is,

$$x_i = \underset{x_i\in X_i}{\operatorname{argmax}}\{v_i(x_i) - b(x_i)\} \qquad (11)$$

The latter objective is not necessarily aligned with the social objective and depends on the billing rule $b(x_i)$. Since the cost function couples the users' variables $x_i, i \in N$, a user's utility depends not only on her/his own profile but also on the other users' consumption choices. This latter fact brings problem (10) in the realm of game theory. In order to bring the system to an equilibrium that optimizes (10), we will draw on the concepts of mechanism design.

We consider a game-theoretic framework, where the ESP announces the billing rule and users iteratively select their preferred allocations, thus formulating the following game $\Gamma$.

*Definition of game $\Gamma$*:
- *Players:* users in $N$
- *Strategies:* each user selects her/his $x_i$, according to (11)
- *Payoffs:* a user's payoff is his/her utility as defined in (5)

Since problem (10) naturally prioritizes users with higher valuation for energy allocation, we need to prevent users from faking a high $v_i(x_i)$. This is the role of the billing rule. The Vickrey-Clarke-Groves (VCG) mechanism has been proven to be the unique welfare-maximizing mechanism that makes it a dominant strategy for each user, to truthfully declare his/her local valuation. Unfortunately, VCG-like mechanisms are not useful here since they violate the *budget-balance* property (*Requirement* 2) and also come with a number of other problems as explained in the introduction. In what follows, we opt for designing a DSM architecture which includes:

a) an indirect and *individually rational* mechanism, including a *budget-balanced* billing rule, implemented in best-response strategies. Although we have to relax the welfare-maximization property, we are actually able to reach a near-optimal solution.

b) an algorithm at the ESP side, which iteratively decides a parameter of the billing rule, thus providing the ESP with online controllability over the system's cost, so that constraint (10a) is satisfied at equilibrium.

## V. PROPOSED DSM ARCHITECTURE

In this section, we present the proposed DSM architecture that fulfills the aforementioned requirements. The presentation is complemented with the presentation of the theorems that prove analytically that the requirements we have set are fulfilled. In more detail, we developed a DSM architecture such that:

a) game $\Gamma$ admits to a Nash Equilibrium (NE)
b) users' actions converge to NE via best-response dynamics
c) the DSM mechanism provides the ESP with controllability over the system's cost, which means that the ESP brings the system to an equilibrium that respects constraint (10a), in case that it is possible.
d) the allocation at equilibrium is as close as possible to the optimal value of problem (10).

### A. The billing rule

Best-response dynamics means that, at each iteration, each user chooses his/her strategy assuming the strategies $x_{-i}$ of other users to be constant. Thus, from a user's perspective, at a certain iteration, his/her bill only depends on his/her own choice of $x_i$. The following equation presents the proposed billing rule:

$$b(x_i) = \sum_{t \in H}\left[x_i^t \cdot \frac{c^t(\sum_{j \in N} x_j^t)}{\sum_{j \in N} x_j^t}\right] + \gamma \cdot \left[\sum_{t \in H}[x_i^t \cdot \sum_{j \neq i}(x_j^t)] - \frac{\sum_{j \in N}[\sum_{t \in H}(x_j^t \cdot \sum_{k \neq j} x_k^t)]}{n}\right] \quad (12)$$

The first term of the sum is identical to existing billing rules. The second term has the purpose to reward/penalize flexibility/inflexibility (ability of user $i$ to modify energy consumption profile). The value of $\gamma$ is iteratively updated by the ESP. The rationale of Eq. (12) is that it penalizes users for synchronizing their loads with others and uses the penalties for rewarding users who counter-balance the aggregated consumption by consuming their load at off-peak timeslots. With respect to the billing rule, we state the following lemma:

*Lemma 2:* For constant values of $x_{-i}$, the bill $b(x_i)$, given by Eq. (12), is strictly convex in $x_i$.

The proof of *Lemma 2* is given in Appendix A. ∎

The user communicates his/her demand profile $x_i$ to the ESP and receives the respective bill $b(x_i)$. Since problem (11) is convex (by *Lemma 1* and *Assumption 3*) the user can apply a gradient projection method to compute his/her best response. Next, we analyze the properties of game $\Gamma$:

*Theorem 1*: A Nash Equilibrium for game $\Gamma$ exists and is unique. Furthermore, best-response dynamics converges to the Nash Equilibrium strategy vector.
The proof of *Theorem 1* is given in Appendix B. ∎

The second term of the sum in (12) introduces a price-discrimination component among users with different levels of flexibility. The ESP can control the magnitude of this discrimination by adjusting parameter $\gamma$, as will be analyzed in subsection *B.* of this section. Thus, by increasing $\gamma$, users are increasingly incentivized to modify their consumption patterns. Note that $\gamma$ does not increase the bills in general but only controls the way that the system's cost is shared among users. This provides an intuition on the way (12) keeps the system *budget balanced*, and is proved formally below.

*Theorem 2*: The billing rule $b(x_i)$, given by Eq. (12), satisfies the *budget balance* property.
The proof is given in Appendix C. ∎

Furthermore, for the proposed billing rule given by Eq. (12), we also verify *Requirement 2*.

*Theorem 3*: Game $\Gamma$, in equilibrium, satisfies the *individual rationality* property.
The proof is given in Appendix D. ∎

### B. The ESP's algorithm for constraint satisfaction

While the users are concerned with maximizing their utility, the ESP is responsible for satisfying constraint (10a). As discussed earlier, the ESP controls the system's cost via parameter $\gamma$. A low value of $\gamma$ would lead to high energy cost, while a large value of $\gamma$ would have a negative impact on the welfare. Thus, the proposed DSM architecture also needs an algorithm for the ESP to identify the appropriate choice of $\gamma$, which brings the system to an allocation that respects constraint (10a) with the least possible sacrifice on the users' utility. Since the ESP is agnostic of the users' valuation functions, determining the appropriate $\gamma$ calls for a global optimization approach. We opt for a Simulated Annealing (SA) method for determining $\gamma$. The entire DSM procedure is depicted in table 1.

**Table 1 The proposed DSM procedure**

| | |
|---|---|
| 1 | set $x_i^t = \overline{x_i}$, $\forall i \in N$, $k = 0, T_0, \gamma_0, \Delta_0$ |
| 2 | **Repeat** |
| 3 |    **Repeat** |
| 4 |       for $i \in N$ |
| 5 |          $i$ calculates best-response $x_i(\gamma_k)$ from (11) |
| 6 |    **until** Nash Equilibrium |
| 7 | ESP calculates cost at NE ($C_k$) |
| 8 | $\Delta_k = (C_k - C_{ref})^2 - (C_{k-1} - C_{ref})^2$ |
| 9 | $k = k + 1$ |
| 10 | $T_k = T_0 \cdot 0.95^k$ |
| 11 | ESP determines next $\gamma_k$ as a function of $T_k, \Delta_{k-1}, \gamma_{k-1}$ |
| 12 | **Until** $\frac{\sum_{k-500}^{k} \Delta_k}{500} \leq \varepsilon$ |

Parameter $T_k$ is the so called "temperature" of the SA

algorithm. In line 11, the SA algorithm determines the next value of $\gamma_k$ based on a probabilistic calculation, which is not presented here.

## VI. APPENDIX

### A) Proof of Lemma 2

We denote by $\mathcal{H}^b$ the Hessian matrix of function $b(\boldsymbol{x}_i)$, defined in Eq. (12). We have to show that $\mathcal{H}$ is positive definite. By substituting Eq. (6) in Eq. (12), we have

$$b(\boldsymbol{x}_i) = \sum_{t \in H}[x_i^t \cdot c \cdot (\sum_{j \in N} x_j^t)] + \gamma \cdot [\sum_{t \in H}[x_i^t \cdot \sum_{j \neq i}(x_j^t)] - \frac{1}{n} \cdot \sum_{j \in N}[\sum_{t \in H}(x_j^t \cdot \sum_{k \neq j} x_k^t)]]$$

$$= \sum_{t \in H}[x_i^t \cdot c \cdot (\sum_{j \in N} x_j^t)] + \gamma \cdot [\frac{n-1}{n} \cdot \sum_{t \in H}[x_i^t \cdot \sum_{j \neq i}(x_j^t)] - \frac{1}{n} \cdot \sum_{j \neq i}[\sum_{t \in H}(x_j^t \cdot \sum_{k \neq j} x_k^t)]]$$

By taking the derivatives:

$$\frac{\partial b(\boldsymbol{x}_i)}{\partial x_i^{t_1}} = c \sum_{j \in N} x_j^{t_1} + c x_i^{t_1} + \gamma \left[\frac{n-1}{n} \sum_{j \neq i}(x_j^{t_1}) - \frac{1}{n} \sum_{j \neq i} x_j^{t_1}\right]$$

and

$$\mathcal{H}^b_{t_1 t_2} = \frac{\partial^2 b(\boldsymbol{x}_i)}{\partial x_i^{t_2} \partial x_i^{t_1}} = \begin{cases} 2c, t_1 = t_2 \\ 0, t_1 \neq t_2 \end{cases}$$

Thus, $\mathcal{H}^b = diag(2c)$ is positive definite.

### B) Proof of Theorem 1

a) The user's payoff is his/her utility given by eq. (5). The first term is concave in $\boldsymbol{x}_i$ by *Assumption 3*. The second term is strictly convex in $\boldsymbol{x}_i$ by *Lemma 1*. Hence, for $x_i^t \geq 0$, $U_i(\boldsymbol{x}_i)$ is strictly concave in $\boldsymbol{x}_i$. Since this holds for every user, we have that $\Gamma$ is a strictly concave n-person game. Thus, by [48, th.1], we have that a NE exists.

b) By [49], it suffices to show that $\Gamma$ is an exact potential game with a concave potential function. Indeed, consider the function:

$$\wp(X) = \sum_{i \in N} v_i(\boldsymbol{x}_i) - \sum_{t \in H} \left\{ \frac{c}{2} \cdot [(\sum_{i \in N} x_i^t)^2 + \sum_{i \in N}[(x_i^t)^2]] + \frac{\gamma(n-2)}{2n} \cdot \sum_{i \in N}(x_i^t \cdot \sum_{j \neq i}(x_j^t)) \right\}$$

Function $\wp(X)$ has the property of potential:
$$\nabla_{\boldsymbol{x}_i} \wp(X) = \nabla_{\boldsymbol{x}_i} U_i(\boldsymbol{x}_i), \quad \forall i \in N$$

Moreover, $\sum_{i \in N} v_i(\boldsymbol{x}_i)$ is concave in $X$ (concave in $\boldsymbol{x}_i$ by *Assumption 3* and zero in $\boldsymbol{x}_j, \forall j \neq i$). Thus, it suffices to prove that the term

$$\wp_2 = -\sum_{t \in H} \left\{ \frac{c}{2} \cdot [(\sum_{i \in N} x_i^t)^2 + \sum_{i \in N}[(x_i^t)^2]] + \frac{\gamma(n-2)}{2n} \cdot \sum_{i \in N}[x_i^t \cdot \sum_{j \neq i}(x_j^t)] \right\}$$

is also concave, or equivalently that $-\wp_2$ is convex.

It is $\nabla_{\boldsymbol{x}_i}(-\wp_2) = \nabla_{\boldsymbol{x}_i} b_i(\boldsymbol{x}_i)$ which yields $\nabla^2_{\boldsymbol{x}_i}(-\wp_2) = \nabla^2_{\boldsymbol{x}_i} b_i(\boldsymbol{x}_i) = \mathcal{H}^b_i$ which by Lemma 1 is positive definite. Hence, $\wp_2$ is also concave in $X$, since its Hessian $\mathcal{H}^{\wp_2} = \nabla^2_X \wp_2 = blkdiag(\{\mathcal{H}^b_i\}_{i=1}^n)$ is a block diagonal positive definite matrix. Hence $\wp$ is concave in $X = (\boldsymbol{x}_1, ..., \boldsymbol{x}_i, ..., \boldsymbol{x}_n)$ as a sum of concave functions in $X$.

c) Since the potential function is concave and players maximize, it directly follows that best-response dynamics converges to the unique NE.

### C) Proof of Theorem 2

It suffices to show that
$$\sum_{i \in N} b(\boldsymbol{x}_i) = \sum_{t \in H} [C^t(\sum_{i \in N} x_i^t)]$$

By substituting $b(\boldsymbol{x}_i)$ from Eq. (12), we have

$$\sum_{i \in N} b(\boldsymbol{x}_i) = \sum_{i \in N} \left\{ \sum_{t \in H} \left[ \frac{x_i^t}{\sum_{j \in N}[x_j^t]} \cdot C^t(\sum_{j \in N} x_j^t) \right] + \gamma \cdot \left[ \sum_{t \in H}[x_i^t \cdot \sum_{j \neq i}(x_j^t)] - \frac{\sum_{j \in N}[\sum_{t \in H}(x_j^t \cdot \sum_{k \neq j} x_k^t)]}{n} \right] \right\}$$

$$= \sum_{t \in T} \sum_{i \in N} \left[ \frac{x_i^t}{\sum_{j \in N}[x_j^t]} \cdot C^t(\sum_{j \in N} x_j^t) \right] + \gamma \sum_{i \in N} \sum_{t \in H}[x_i^t \cdot \sum_{j \neq i}(x_j^t)] - \gamma \cdot \sum_{i \in N} \left[ \frac{\sum_{j \in N}[\sum_{t \in H}(x_j^t \cdot \sum_{k \neq j} x_k^t)]}{n} \right]$$

$$= \sum_{t \in T} [C^t(\sum_{i \in N} x_i^t)] + \gamma \cdot [\sum_{i \in N} \sum_{t \in H}(x_i^t \cdot \sum_{j \neq i}(x_j^t)) - \sum_{j \in N} \sum_{t \in H}(x_j^t \cdot \sum_{k \neq j}(x_k^t))] = \sum_{t \in T} [C^t(\sum_{i \in N} x_i^t)]$$

which completes the proof.

### D) Proof of Theorem 3

The root vector $\boldsymbol{x}_i^{root} \triangleq \{x_i^{t,root}\}, t \in H$, for which $b(\boldsymbol{x}_i^{root}) = 0$, is derived by solving from Eq. (12):

$$x_i^{t,root} = \frac{1}{2cn} \left[ -\sum_{j \neq i}(x_j^t) \cdot (n(c+1) - \gamma - 1) \pm \sqrt{[\sum_{j \neq i}(x_j^t) \cdot (n(c+1) - \gamma - 1)]^2 + 4c\gamma n} \right]$$

Setting $\sum_{j \neq i} x_j^t \cdot (n(c+1) - \gamma - 1) = \alpha$ and $4c\gamma n = \beta$, we get

$$x_i^{t,root} = \frac{1}{2cn}[-\alpha \pm \sqrt{\alpha^2 + \beta}]$$

which means that there is always exactly one $x_i^{t,root} \geq 0$. By *Assumptions* 1 and 2, we have $v_i(x_i^{t,root}) \geq 0$. Thus, from Eq. (5) we get that $U_i(\boldsymbol{x}_i^{root}) \geq 0$. This means that each user's utility is weakly positive, which completes the proof.

## VII. REFERENCES

[1] L. Gkatzikis, I. Koutsopoulos, and T. Salonidis, "The role of aggregators in smart grid demand response markets," IEEE J. Sel. Areas Commun., vol. 31, no. 7, pp. 1247–1257, Jul. 2013.
[2] SOCIALENERGY: A Gaming and Social Network Platform for Evolving Energy Markets' Operation and Educating Virtual Energy Communities, D6.1 Data Management Plan, dissemination and exploitation plans. Available Online: http://socialenergy-project.eu/
[3] A. C. Chapman, G. Verbic, and D. J. Hill, "A healthy dose of reality for game-theoretic approaches to residential demand response," in Proc. Bulk Power Syst. Dyn. Control IX Optim. Security Control Emerg. Power Grid (IREP), Rethymno, Greece, 2013, pp. 1–13.
[4] P. Samadi, A. H. Mohsenian-Rad, R. Schober, V. W. S. Wong, "Advanced demand side management for the future smart grid using mechanism design", IEEE Trans. Smart Grid, vol. 3, no. 3, pp. 1170-1180, Sep. 2012.
[5] RESCOOP: European federation of renewable energy cooperatives. Available Online: https://www.rescoop.eu/
[6] ECOPOWER, Available Online: http://www.ecopower.com/
[7] VIMSEN: Virtual Micro grids for smart energy networks, Available Online: http://www.ict-vimsen.eu/
[8] M. Stadler, G. Cardoso, S. Mashayekh, T. Forget, N. DeForest, A. Agarwal, A. Schönbein, "Value streams in microgrids: A literature review", Appl. Energy, vol. 162, pp. 980-989, Oct. 2016.
[9] S. Borenstein, "Understanding competitive pricing and market power in wholesale electricity markets", The Elec. J., vol. 13, pp. 49-57, 2000
[10] USEF: A flexibility market, Available Online: https://www.usef.energy/flexibility/
[11] Community energy storage DNV-GL, Available Online: https://www.dnvgl.com/energy/index.html

[12] DIRECTIVE 2012/27/EU OF THE EUROPEAN PARLIAMENT AND OF THE COUNCIL, Official Journal of the European Union, L 315/1, Oct 2012. Available Online: https://ec.europa.eu/energy/en/topics/energy-efficiency/energy-efficiency-directive

[13] Policy development for improving RES-H/C penetration in European Member States. Available Online: https://ec.europa.eu/energy/intelligent/projects/en/projects/res-h-policy

[14] N. Li, J. Marden, "Decoupling coupled constraints through utility design", IEEE Trans. Autom. Control, vol. 59, no. 8, pp. 2289-2294, Aug. 2014

[15] P. Makris, N. Efthymiopoulos, V. Nikolopoulos, A. Pomazanskyi, B. Irmscher, K. Stefanov, K. Pancheva, E. Varvarigos, "Digitization Era for Electric Utilities: A Novel Business Model Through an Inter-Disciplinary S/W Platform and Open Research Challenges", IEEE Access, vol.6, pp. 22452-22463, 2018.

[16] K. Steriotis, G. Tsaousoglou, N. Efthymiopoulos, P. Makris, E. Varvarigos, "A novel behavioral real time pricing scheme for the active energy consumers' participation in emerging flexibility markets", Sust. Energy, Grids and Networks (SEGAN), vol.16, pp. 14-27, Dec. 2018.

[17] G. Tsaousoglou, N. Efthymiopoulos, P. Makris, E. Varvarigos "Personalized real time pricing for efficient and fair demand response in energy cooperatives and highly competitive flexibility markets" to be published, DOI :10.1007/s40565-018-0426-0

[18] K. Steriotis, G. Tsaousoglou, N. Efthymiopoulos, P. Makris and E. Varvarigos, "Development of real time energy pricing schemes that incentivize behavioral changes," 2018 IEEE International Energy Conference (ENERGYCON), Limassol, Cyprus, pp. 1-6, 2018.

[19] A. C. Chapman and G. Verbič, "An Iterative On-Line Auction Mechanism for Aggregated Demand-Side Participation," in IEEE Trans. Smart Grid, vol. 8, no. 1, pp. 158-168, Jan. 2017.

[20] G. Tsaousoglou, P. Makris, E. Varvarigos, " Electricity market policies for penalizing volatility and scheduling strategies: The value of aggregation, flexibility, and correlation", Sust. Energy, Grids and Networks (SEGAN), vol.12, pp. 57-68, Dec. 2017.

[21] G. Tsaousoglou, P. Makris and E. Varvarigos, "Micro grid scheduling policies, forecasting errors, and cooperation based on production correlation," 2016 2nd International Conference on Intelligent Green Building and Smart Grid (IGBSG), Prague, pp. 1-6, 2016.

[22] S. Mhanna, G. Verbič and A. C. Chapman, "A Faithful Distributed Mechanism for Demand Response Aggregation," in IEEE Trans. Smart Grid, vol. 7, no. 3, pp. 1743-1753, May 2016.

[23] M. H. K. Tushar, C. Assi and M. Maier, "Distributed Real-Time Electricity Allocation Mechanism for Large Residential Microgrid," in IEEE Trans. Smart Grid, vol. 6, no. 3, pp. 1353-1363, May 2015.

[24] C. Chen, J. Wang, Y. Heo and S. Kishore, "MPC-Based Appliance Scheduling for Residential Building Energy Management Controller," in IEEE Trans. Smart Grid, vol. 4, no. 3, pp. 1401-1410, Sept. 2013.

[25] H. M. Soliman and A. Leon-Garcia, "Game-Theoretic Demand-Side Management With Storage Devices for the Future Smart Grid," in IEEE Trans. Smart Grid, vol. 5, no. 3, pp. 1475-1485, May 2014.

[26] S. Bahrami and M. Parniani, "Game Theoretic Based Charging Strategy for Plug-in Hybrid Electric Vehicles," in IEEE Trans. Smart Grid, vol. 5, no. 5, pp. 2368-2375, Sept. 2014.

[27] J. Ma, J. Deng, L. Song and Z. Han, "Incentive Mechanism for Demand Side Management in Smart Grid Using Auction," in IEEE Trans. Smart Grid, vol. 5, no. 3, pp. 1379-1388, May 2014.

[28] Z. Baharlouei, H. Narimani and M. Hashemi, "On the Convergence Properties of Autonomous Demand Side Management Algorithms," in IEEE Trans. Smart Grid.

[29] Z. Zhao, W. C. Lee, Y. Shin and K. B. Song, "An Optimal Power Scheduling Method for Demand Response in Home Energy Management System," in IEEE Trans. Smart Grid, vol. 4, no. 3, pp. 1391-1400, Sept. 2013.

[30] P. Samadi, H. Mohsenian-Rad, V. W. S. Wong and R. Schober, "Tackling the Load Uncertainty Challenges for Energy Consumption Scheduling in Smart Grid," in IEEE Trans. Smart Grid, vol. 4, no. 2, pp. 1007-1016, June 2013.

[31] A. H. Mohsenian-Rad, V. W. S. Wong, J. Jatskevich, R. Schober and A. Leon-Garcia, "Autonomous Demand-Side Management Based on Game-Theoretic Energy Consumption Scheduling for the Future Smart Grid," in IEEE Trans. Smart Grid, vol. 1, no. 3, pp. 320-331, Dec. 2010.

[32] Z. Baharlouei, M. Hashemi, H. Narimani and H. Mohsenian-Rad, "Achieving Optimality and Fairness in Autonomous Demand Response: Benchmarks and Billing Mechanisms," in IEEE Trans. Smart Grid, vol. 4, no. 2, pp. 968-975, June 2013.

[33] N. Yaagoubi and H. T. Mouftah, "User-Aware Game Theoretic Approach for Demand Management," in IEEE Trans. Smart Grid, vol. 6, no. 2, pp. 716-725, March 2015.

[34] Z. Wang and R. Paranjape, "Optimal Residential Demand Response for Multiple Heterogeneous Homes With Real-Time Price Prediction in a Multiagent Framework," in IEEE Trans. Smart Grid, vol. 8, no. 3, pp. 1173-1184, May 2017.

[35] L. P. Qian, Y. J. A. Zhang, J. Huang and Y. Wu, "Demand Response Management via Real-Time Electricity Price Control in Smart Grids," in IEEE Journal on Selected Areas in Communications, vol. 31, no. 7, pp. 1268-1280, July 2013.

[36] N. Li, L. Chen and S. H. Low, "Optimal demand response based on utility maximization in power networks" 2011 IEEE Power and Energy Society General Meeting, Detroit, MI, USA, pp. 1-8, 2011.

[37] A. H. Mohsenian-Rad and A. Leon-Garcia, "Optimal Residential Load Control With Price Prediction in Real-Time Electricity Pricing Environments," in IEEE Trans. Smart Grid, vol. 1, no. 2, pp. 120-133, Sept. 2010.

[38] P. Samadi, A. H. Mohsenian-Rad, R. Schober, V. W. S. Wong and J. Jatskevich, "Optimal Real-Time Pricing Algorithm Based on Utility Maximization for Smart Grid," 2010 First IEEE International Conference on Smart Grid Communications, Gaithersburg, MD, pp. 415-420, 2010.

[39] R. Deng, Z. Yang, J. Chen, N. R. Asr and M. Y. Chow, "Residential Energy Consumption Scheduling: A Coupled-Constraint Game Approach," in IEEE Trans. Smart Grid, vol. 5, no. 3, pp. 1340-1350, May 2014.

[40] E. Nekouei, T. Alpcan and D. Chattopadhyay, "Game-Theoretic Frameworks for Demand Response in Electricity Markets," in IEEE Trans. Smart Grid, vol. 6, no. 2, pp. 748-758, March 2015.

[41] R. Johari, "The Price of Anarchy and the Design of Scalable Resource Allocation Mechanisms" in Algorithmic Game Theory, Cambridge University Press, pp 543-568, 2007.

[42] S. Althaher, P. Mancarella and J. Mutale, "Automated Demand Response From Home Energy Management System Under Dynamic Pricing and Power and Comfort Constraints," in IEEE Trans. Smart Grid, vol. 6, no. 4, pp. 1874-1883, July 2015.

[43] X. Chen, T. Wei and S. Hu, "Uncertainty-Aware Household Appliance Scheduling Considering Dynamic Electricity Pricing in Smart Home," in IEEE Trans. Smart Grid, vol. 4, no. 2, pp. 932-941, June 2013.

[44] O. Erdinç, A. Taşcıkaraoğlu, N. G. Paterakis, Y. Eren and J. P. S. Catalão, "End-User Comfort Oriented Day-Ahead Planning for Responsive Residential HVAC Demand Aggregation Considering Weather Forecasts," in IEEE Trans. Smart Grid, vol. 8, no. 1, pp. 362-372, Jan. 2017

[45] W. Tang, S. Bi and Y. J. Zhang, "Online Coordinated Charging Decision Algorithm for Electric Vehicles Without Future Information," in IEEE Trans. Smart Grid, vol. 5, no. 6, pp. 2810-2824, Nov. 2014.

[46] J.M. Perlof "Microeconomics", 7th Edition, Pearson, 2015.

[47] D. P. Kothari, I. J. Nagrath, "Modern Power System Analysis", McGraw-Hill, 2003.

[48] J. B. Rosen, "Existence and Uniqueness of Equilibrium Points for Concave N-Person Games", in Econometrica, vol. 33, No. 3, pp. 520-534, Jul. 1965.

[49] D. Monderer and L. S. Shapley, "Potential games", Games and economic behavior, vol. 14, no. 1, pp. 124–143, 1996.